\documentclass[%
superscriptaddress,
preprint,
amsmath,amssymb,
aip,
]{revtex4-2}

\usepackage{graphicx}
\usepackage{dcolumn}
\usepackage{bm}
\usepackage[mathlines]{lineno}
\usepackage{lineno}

\makeatletter
\let\@fnsymbol\@fnsymbol@latex
\@booleanfalse\altaffilletter@sw
\makeatother

\usepackage{xcolor}
\usepackage{ulem}

\begin{document}

\title{Nonlinear phononics in 2D SnTe: a ferroelectric material with phonon dynamical amplification of electric polarization}


\author{Dongbin Shin}
\affiliation{Max Planck Institute for the Structure and Dynamics of Matter and Center for Free-Electron Laser Science, Luruper Chaussee 149, 22761, Hamburg, Germany}

\author{Shunsuke A. Sato}
\affiliation{Max Planck Institute for the Structure and Dynamics of Matter and Center for Free-Electron Laser Science, Luruper Chaussee 149, 22761, Hamburg, Germany}
\affiliation{Center for Computational Sciences, University of Tsukuba, Tsukuba 305-8577, Japan}

\author{Hannes H{\"u}bener}
\affiliation{Max Planck Institute for the Structure and Dynamics of Matter and Center for Free-Electron Laser Science, Luruper Chaussee 149, 22761, Hamburg, Germany}
\author{Umberto De Giovannini}
\affiliation{Max Planck Institute for the Structure and Dynamics of Matter and Center for Free-Electron Laser Science, Luruper Chaussee 149, 22761, Hamburg, Germany}
\affiliation{Nano-Bio Spectroscopy Group, Departamento de Fsica de Materiales, Universidad del Pas Vasco, 20018 San Sebastian, Spain}
\author{Noejung Park}\email[]{noejung@unist.ac.kr}
\affiliation{Department of Physics, Ulsan National Institute of Science and Technology, UNIST-gil 50, Ulsan 44919, Korea}
\author{Angel Rubio}\email[]{angel.rubio@mpsd.mpg.de}
\affiliation{Max Planck Institute for the Structure and Dynamics of Matter and Center for Free-Electron Laser Science, Luruper Chaussee 149, 22761, Hamburg, Germany}
\affiliation{Nano-Bio Spectroscopy Group, Departamento de Fsica de Materiales, Universidad del Pas Vasco, 20018 San Sebastian, Spain}
\affiliation{Center for Computational Quantum Physics (CCQ), The Flatiron Institute, 162 Fifth avenue, New York NY 10010}

\begin{abstract}
Ultrafast optical control of ferroelectricity using intense terahertz fields has attracted significant interest. 
Here we show that the nonlinear interactions between two optical phonons in SnTe, a two-dimensional in-plane ferroelectric material, enables a dynamical amplification of the electric polarization within subpicoseconds time domain. 
Our first principles time dependent simulations show that the infrared-active out-of-plane phonon mode, pumped to nonlinear regimes, spontaneously generates in-plane motions, leading to rectified oscillations in the in-plane electric polarization. 
We suggest that this dynamical control of ferroelectric material, by nonlinear phonon excitation, can be utilized to achieve ultrafast control of the photovoltaic or other non-linear optical responses.
\end{abstract}

\maketitle

\section*{Keywords} Nonlinear phononics, Feroelectricity, photovoltaics

\section*{Introduction}

Light-matter interaction has become the main gateway through which the microscopic quantum materials properties are interacted, and hence engineered, with tools of laboratories in the macroscopic world.\cite{Hillenbrand2002,Britnell2013,Nova2019,Li2019}
Optical responses of non-centrosymmetric insulators are paradigmatic examples in this context.\cite{Tokura2018}
They can host various photo-electric dynamical phenomena, such as shift current, second harmonic generation and circular photo-galvanic effect.\cite{J.E.Sipe2000,DeJuan2017,Xu2018,Young2012}
Moreover, various driven states induced by intense low-frequency external fields attracted substantial interest within various different context in recent years.\cite{Sodemann2015,Ma2019}
For example, the nonlinear Hall current in inversion-asymmetric materials shows a second order responses to low-frequency driving fields, controlled by asymmetric distribution of Berry curvature (more specifically, the Berry curvature dipole).\cite {Sodemann2015}

Defining characteristics of ferroelectricity is the existence of switchable bi-stable configurations, which share the same space group but with opposite electric polarization.\cite{Fei2016,Chang2016}
However, recent studies using optical pump, in terahertz or low infra-red band, have expanded the scope of ferroelectricity to the non-equilibrium regime of driven states.\cite{Nova2019,Li2019,Fu2000,Catalan2011}
In this field, low-dimensional materials or nano-structured systems are more attractive in view of few-atom scale local switching that can lead to even enhanced non-volatility of state.\cite{Ma2019,Chang2016,Fei2016,Kim2019,Nova2019,Fu2000,Catalan2011,Rubio-Marcos2015,Li2019} 
An exemplary system in this area is the monolayer SnTe: the in-plane ferroelectricity of the atom-thick layer has Curie temperature of 270K, three-times higher than that of the corresponding three-dimensional ferroelectric bulk structure (100K).\cite{Chang2016}

In this study, we investigate the interacting phonon dynamics in this ferroelectric SnTe monolayer. 
Through real-time $ab~initio$ molecular dynamics simulations, we find that the coherently pumped out-of-plane $A_u$ phonon mode, up to nonlinear regimes, spontaneously entails the in-plane $E_{u,x}$ vibrations. 
Remarkably, the entangled phonon dynamics induces the in-plane polarization to increase over time in a rectified manner. 
This non-linear phonon interaction is explained in terms of the modification of the potential energy surface by the phonon modes, and the microscopic contributions to the polarization changes are analyzed on the standpoint of the electronic part of the Born effective charge. 
We discuss that this dynamically amplified ferroelectricity makes the variations of the Berry curvature field more steep, which results in enhanced Berry curvature dipole and also the increased stability of the ferroelectric phase.

\section*{Results}
\subsection*{Atomic geometry and electronic structure of monolayer SnTe}
Monolayer SnTe consists of alternating Sn and Te atoms in a puckered rectangular lattice, as shown in Fig.~1a. 
Previous experimental studies revealed that the in-plane ferroelectricity of monolayer SnTe originates from the staggered atomic displacement between Sn and Te, which otherwise comprises the cubic rock-salt structure.\cite{Chang2016}
In the ferroelectric configuration, the inversion symmetry is lifted along with the removal of the mirror plane perpendicular to the $x$-axis while the mirror plane along the $y$-axis ($M_y$) is well preserved, as depicted by the dashed line in Fig.~1a.
This equilibrium structure (one of the energy minima shown in Fig.~1b) is to be characterized by the space group of $Pmn2_1$.\cite{Chang2016,Sawinska2020,Kim2019}
To visualize the double-well potential energy with respect to the in-plane polarization, as depicted in Fig.~1b, the atomic structures are linearly interpolated (or extrapolated) from the two equilibrium geometries at each equilibrium valley.
The optimized geometry of monolayer SnTe, obtained by minimizing the total energies of density functional theory (DFT), is distorted by $70m$\AA ~along the $x$-axis as shown in Fig.~1a, producing the in-plane polarization of $P_x=\pm13.9 \times  10^{-12} Cm^{-1}$.\cite{Kim2019}
The monolayer SnTe is a seminconductor with $0.6$ eV indirect band gap, which is underestimated value by DFT calculation comparing with experimental gap ($1.6$eV),\cite{Chang2016,Xu2017} between the valence band maximum at a point on the  $\Gamma$X line, and conduction band minimum at a point on $\Gamma$Y, as indicated by upward and downward arrows in Fig.~1c, respectively.
This geometric nature of the monolayer SnTe is well reflected in the distribution of the Berry curvature which is asymmetric with respect to the same mirror plane ($M_y$), as indicated by the dashed line in Fig.~1d. 
Previous studies have suggested that this asymmetric Berry curvature can be attributed to the orbital Rashba effect.\cite{Kim2019}
Upon hole doping, this asymmetry in the Berry curvature grants sizable Berry curvature dipole ($\Lambda_y$) in the in the $y$-direction which can be manifested helicity-dependent second-order electro-optical responses.\cite{Sodemann2015,Ma2019}

\subsection*{Non-linear lattice dynamics and electric polarization}
We now show how the in-plane polarization is delicately intertwined with selected phonon dynamics.
We first briefly describe the phonon structures of the monolayer SnTe, as depicted in Fig.~2a.\cite{Li20192}
The optical phonons can be classified into three groups: the in-plane oscillations at around $1.2$ to $1.6$ THz, the out-of-plane modes at higher frequency range from $4.0$ to $4.8$ THz, and the intermediate bands around $3.0$ THz that comes form the mixed motions. 
In the present work, we focus on the phonons that can be directly coupled with the electronic dipole oscillation. 
In this monolayer SnTe, the in-plane motions ($E_{u,x}$ and $E_{u,y}$) and the out-of-plane $A_u$ mode are active for infra-red field (IR-active).
The $A_u$ phonon mode mainly consists of the alternating atomic motion between the Sn and Te in the out-of-plane direction with the period of $216$~fs period as shown in Fig.~2a~(upper). 
The $E_{u,x}$ phonon is characterized by the alternating motion between Sn and Te along the $x$-direction, has a period of $742$~fs, as shown in Fig.~2a~(below).
The details of phonon dispersion and LO-TO splitting of $E_u$ is discussed in Supplementary Note 1 and Supplementary Figure 1.

To quantitatively gauge the effect of the coherently pumped phonons on the electronic structure, we performed the DFT Born-Oppenheimer (BO) molecular dynamics starting from the ground state atomic positions displaced along the eigenvector of the $A_u$ phonon mode ($Q_{A_u}^{init}=80$~m\AA). 
Figure~2b shows that the obtained $z$-direction motions in the BO dynamics indeed prove the $A_u$ mode vibration with the frequency of $216$~fs. 
On the different instantaneous atomic configuration along this lattice dynamics, we calculated the electric polarization, of which the in-plane component is shown in Fig.~2c. 
Unexpectedly, instead of vibrating back and forth around the equilibrium point, the induced in-plane polarization oscillates in a rectified manner leading to increased polarizations at all time. Furthermore, this polarization oscillation complies with the frequency of the in-plane $E_{u,x}$ phonon mode ($742$~fs) rather than that of the driving of $A_u$ phonon mode ($216$~fs).
This dynamical polarization is mainly contributed by the electronic part, while the change in the ionic contribution is negligible.
Similarly, the atomic displacements in the $x$-direction, recorded in Fig. 2d, mainly follows the pattern of $E_{u,x}$ phonon: the major oscillation reveal the period of $742$~fs, while the minor secondary oscillations with higher frequencies are deemed to reflect the beating between the two phonon modes. 
However, it is worth noting that this induced in-plane vibration deviates from the static equilibrium point. 
The oscillation centers for two Sn atoms shifted positively (indicated by the upward arrow in Fig.~2d), whereas that of two Te atoms are relocated negatively (indicated by the negative arrow in Fig.~2d), constituting the increased dipole moment.
In short, our results indicate that a pumping of the out-of-plane IR-active $A_u$ mode ($4.6$ THz) over a non-linear regime derives the lower-frequency in-plane vibration of $E_{u,x}$ mode, which in turn gives rise to the in-plane oscillation of the polarization near terahertz ($1.3$ THz).

To analyze in detail the effect of each phonon mode on the electric polarization, we distorted the equilibrium lattice along each of the phonon eigenvectors, and then calculated the electric polarization. 
As summarized in the Fig.~3a, the variation in the polarization scales almost linearly with the atomic movement in the $E_{u,x}$ mode. 
In contrast, the atomic displacement along the out-of-plane $A_u$ mode has no apparent effect in the shown window in the upper panel of Fig.~3a. 
However, when the displacement is further increased, as shown in the bottom panel of Fig.~3a, a trace of a parabolic dependence emerges, which should be attributed to the interaction of the $E_{u,x}$ mode into the $A_u$ mode in the non-linear range.
It is worth noting that, as shown in the bottom panel of Fig.~3a, both the positive and negative displacement in the $A_u$ mode increases the in-plane polarization, irrespective of the phases of $A_u$ phonon oscillations, leading to the rectified oscillations. 

To access the non-linear interaction between the $E_{u,x}$ and $A_u$ phonon modes, we calculated the total energy of the lattice with the additional displacement along the $E_{u,x}$ mode on top of a given distortion along the $A_u$ mode: the potential energy surface along the additional $E_{u,x}$ coordinate hereafter is denoted as $\Delta E_{Q_{A_u}}[Q_{E_{u,x}}]$, which is plotted in Fig.~3b.
Notice that the potential energy shown in the $E_{u,x}$ coordinate is plotted with respect to the minimum energy of configurations with given $Q_{A_u}$.
Irrespective of whether the movement in the $A_u$ mode is positive or negative ($Q_{A_u}=+80m$\AA~and $Q_{A_u}=-80m$\AA~in Fig.~3b), the energy minimum shifts positively in the $E_{u,x}$ coordinate (as indicated by the positive arrow in Fig.~3b).
This can be explained by considering the Coulomb interaction between the asymmetric ionic chain in the primitive cell.
At the equilibrium, the Sn and Te atoms alternates in the $x$-axis with small in-plane displacement, as indicated by $x$ in the cross-sectional view, shown in Fig.~3c. 
If the atomic positions are displaced toward the $A_u$ phonon mode, depicted by $d/2$ in the middle panel of Fig.~3c, the bond distances between the pair of Sn and Te become $b_{\mp} =\sqrt{x^2+(z\mp d)^2}$ ($b_-$ for left pair (Sn1-Te1) and $b_+$ for right pair (Sn2-Te2) in Fig.~3c), which provide repulsive and attractive forces between Sn and Te atoms, respectively.
The repulsive restoring force on the compressed bond is depicted in Fig.~3c, and the attractive restoring force is expected from the elongated part. 
The stronger repulsive force produces the in-plane translations of Sn and Te atom from their equilibrium position, resulting in an increased in-plane displacement ($Q_{E_{u,x}}$).
As depicted in the middle and bottom parts of Fig.~3c, the cases of positive and negative $Q_{A_u}$ correspond to mere interchange between the bond pair of the elongated and the compressed: notice the bonds denoted by Sn1-Te1 and Sn2-Te2. 
As a result, the atomic displacement along the $A_u$ phonon mode always increases the in-plane distortions irrespective of whether the $Q_{A_u}$ is positive or negative.

On the other hand, the presence of the mirror plane along $y$-direction ($M_y$), prohibits the in-plane atomic distortion in $y$-direction. 
In other words, $E_{u,y}$ phonons would never be excited by the $A_u$ vibration, once the lattice already has ferroelectric distortion in the $x$-direction.  
This is consistent with the trend shown in Fig.~3a, which reveals that both the positive and negative displacement along the $A_u$ mode derives the positive-direction distortion in the $E_{u,x}$ mode eigenvector. 
Eventually, the vibration in the $A_u$ mode intensifies the in-plane polarization over all time, regardless of the phases in the phononic oscillation.
We explicitly present the double well potential energy surface with respect to the in-plane electric polarization, as shown in Fig.~3d. 
Besides the increase in the in-plane polarization, the barrier of the double well is substantially increased ($\Delta E=$0.8meV) stabilizing the ferroelectric phase. The full range plot of the double well potential energy is discussed in Supplementary Note 2 and Supplementary Figure 2.
The dynamical amplification induced by nonlinear phonon interaction can be compared to the polarization rotation discussed numerously in previous literatures. However, since equilibrium geometry has negligible polarization along y- and z-directions, this dynamic increase in the polarization should be considered as a distinct mechanism.\cite{Fu2000,Catalan2011}

We now quantitatively estimate the effect of this dynamical polarization enhancement on the stability of the ferroelectric phase.
The Curie temperature for ferroelectric transitions in two- and three- dimensional structures, such as monolayer group-IV monochalcogenides and perovskite transition metal oxides, have been very accurately predicted from the double well potential energy surface by using forth-order Landau theory.\cite{Wojdel2014,Fei2016,Grinberg2004}
The free energy ($F$) can be written in terms of spontaneous electric polarization ($P$) with positive constants $\alpha$ and $\beta$ :
\begin{equation}
F=\alpha (T-T_c)P^2 +\beta P^4.
\end{equation}
At the equilibrium polarization, the free energy minimization condition ($dF/dP=0$) requires the Curie temperature to be ${\text T_c}=\frac{2\beta}{\alpha}P^2$. 
By fitting the free energy form to the double well potential given in Fig.~1b, we obtained the constants $\frac{2\beta}{\alpha}=0.967 eV/(10^{-10}Cm^{-1})^2$, and thus we have ${\text T_c}=287K$, which is well matched with experimental observation ($\sim270K$).\cite{Chang2016}
Notice that, while thermal average phonons make the lattice vibrates with respect to the
equilibrium, the coherently pumped $A_u$ phonons shifts the center of vibration, granting an enhanced
in-plane polarization onto lattice over all time. 
On average this rapid oscillation can be interpreted as a static increase in the in-plane polarization. 
With $Q_{A_u}=80m$\AA~ the average increase in the polarization amounts to $15.9 \times 10^{-12}Cm^{-1}$ which renders the Curie temperature ${\text T_c}=358K$ with $\frac{2\beta}{\alpha}=1.213 eV/(10^{-10}Cm^{-1})^2$ according to the fourth-order Landau theory.

\subsection*{Charge oscillation and photovoltaic response functions}
The time variation of the in-plane polarization, as presented in Fig.~2c, indicates that response of the material produces a dipole oscillation with a substantial x-component.  
To quantify these electronic responses within the full $ab~initio$ way, we performed real-time time-propagation of the Kohn-Sham electronic states with the atomic positions being translated concurrently through the Ehrenfest dynamics, as follows: 
\begin{equation}
i\hbar \frac{\partial}{\partial t} |\psi_{n,\bold{k}}(t)>=\hat{H}[\bold{R}(t),\rho(t)] |\psi_{n,\bold{k}}(t)>
\end{equation}
Assuming an impulsive kick, the lattice was initially distorted into an $A_u$ mode eigenvector ($Q_{A_u}^{init}=40m$\AA), and the electronic states were prepared in its ground state at $t=0$. 
Detailed techniques related to the construction of the time-propagation unitary operators are well described elsewhere, including our previous works.\cite{Shin2016,Shin2019} 
The obtained Ehrenfest real-time atomic trajectories show almost identical nonlinear phonon interaction to those of BO molecular dynamics as shown in Supplementary Figure 3.
This indicates that the electronic states do not much deviate from the instantaneous ground states, and non-adiabatic effects are negligible even in the regime of nonlinear phononic oscillations. 
The movement of charges were evaluated from the electric current, as computed with the time-evolving Kohn-Sham states: $\bold{J}(t)=\frac{-e}{m}\sum_{n,\bold{k}} ^{occ} \langle\psi_{n,\bold{k}}(t)|\hat{\bold{p}}|\psi_{n,\bold{k}}(t)\rangle$.\cite{Shin2019}
The real-time profile of the out-of-plane and the in-plane current ($J_z$ and $J_x$) are presented in Figs.~4a and 4b, respectively. 
The corresponding Fourier components are presented in Figs.~4c and 4d. 
On this initial impulsive kick into the $A_u$ phonon eigenvector, the out-of-plane oscillation ($J_z(t)$) is dominated by the $\omega_{A_u}$ component with a sizable band width (Fig. 4c). 
On the other hand, the $J_x(\omega)$ exhibits an apparent $\omega_{E_{u,x}}$ peak, together with $2\omega_{A_u}$ signal, as shown in Fig.~4d. 
The apparent $E_{u,x}$ excitation directly evidences the nonlinear coupling of the $E_{u,x}$ mode to the $A_u$ mode, as explained above.
The mechanism that underlies the frequency doubling ($2\omega_{A_u}$) can be ascribed to the parabolic dependence of the polarization on the $A_u$ mode displacement, as shown in the bottom panel of Fig.~3a: on the harmonic vibration in the $A_u$ mode ($Q_{A_u}(t)\sim sin(\omega_{A_u}t)$), the parabolic polarization should convey the second harmonic responses in the current from $J_x(t) =dP_x^{elec}/dt\sim dQ_{A_u}^2/dt \sim sin(2\omega_{A_u}t)$.
This nonlinear phononic interaction can be detected from the radiation field by tracing the in-plane component of the polarity in-plane polarity, as schematically sketched in left panel of Fig.~4e. 
In realistic experiment, the perfect impulsive kick is hard to realize, but a pulse with a substantial band width centered at $\omega_{A_u}$ can be applied.
If the pulse is sufficiently short, which band width covers the range between $\omega_{E_{u,x}}$ and $\omega_{A_u}$ as depicted in the Fig.~4e, the proposed phenomena can be experimentally observed.

The dynamical amplification of the polarization also has a substantial effect on the photovoltaic optical responses.
Among various characteristics of non-centrosymmetric materials, such as second harmonic generation and circular photogalvanic current, which sharply depends on the polarization, here we examine the effect of the nonlinear phononics on the shift current.\cite{Tokura2018,J.E.Sipe2000,DeJuan2017,Xu2018,Young2012}
The response function for shift current can be evaluated as follows; 
\begin{equation}
\sigma^{abb}(\omega)=\frac{\pi e^3}{\hbar^2}\int \frac{d\bold{k}}{(2\pi)^3}\sum_{nm}f_{nm}(\bold{k})R^a_{nm}(\bold{k})|r^b_{nm}(\bold{k})|\delta(\omega_{nm}-\omega),
\end{equation}
where $f_{nm}$, $R^a_{nm}$, $r^b_{nm}$ and $\omega_{nm}$ are occupation difference, shift vector, Berry connection, and energy difference between $n$-th and $m$-th states, respectively.\cite{J.E.Sipe2000}
The shift vector, which describes the charge center difference between occupied and unoccupied bands, is directly proportional to the internal polarization,\cite{J.E.Sipe2000,Cook2017,Young2012} and thus will be most directly affected by this nonlinear phononics of the $A_u$ phonon. 
As an example, the shift current response function $\sigma^{xxx}(\omega)$, the $x$-direction photoconductivity in response to the incoming light with $x$-direction polarization, is summarized in Fig.~4f with respect to various distortions into the $A_u$ mode.
The responses to a particular frequency of incoming light ($\omega=0.6eV$) and the maximum of $\sigma^{xxx}$ over frequencies are presented in Fig.~4f.
This enhanced shift current is attributed to the increase in the in-plane polarization, which is induced by the nonlinear phononics and thus irrespective of whether the $A_u$ phonon is positively or negatively distorted (see Fig.~2c).
Whether the bulk photovoltaic effect of solid can be altered by phonons has been questioned previously,\cite{Gong2018} and our finding of the rectified oscillation of the polarization and the amplified photovoltaic responses can be considered as a new example in this light of search.

\subsection*{Interaction strength and the effect of hole doping}
The larger initial pumping along the $A_u$ mode produces the stronger entailing in-plane motion in the $E_{u,x}$ mode. 
Here, we quantify the non-linear coupling strength between the two optical phonons and analyze the electronic origin of the coupling in terms of mode effective charge. 
The interaction between these two optical phonons can be described through a nonlinear harmonic oscillator coupling model, as treated in previous literature:\cite{Cartella2017,Fei2016}
\begin{equation}
H=\frac{1}{2}M\dot{Q}_{E_{u,x}}^2+\frac{1}{2}M\dot{Q}_{A_{u}}^2+\frac{1}{2}M\omega^2_{A_u}Q_{A_{u}}^2+\frac{1}{2}M\omega^2_{E_{u,x}}(Q_{E_{u,x}}+g_0 Q_{A_{u}}^2+g_1 Q_{A_{u}}^4)^2,
\end{equation}
Here, $M$ and $g$ are effective mass and the coupling constant, respectively. 
The equations of motion for $Q_{E_{u,x}}(t)$ and $Q_{A_u}(t)$ can be derived from the Hamiltonian dynamics, and their time series can be integrated through the Verlet algorithm. 
As we have assumed an impulsive kick in the $ab~initio$ dynamics study above, the time trajectories were evaluated from various initial displacements ($Q_{A_u}^{init}$) with zero initial velocity. 
By fitting the obtained $Q_{E_{u,x}}(t)$ to the BO molecular dynamics results, shown in Fig.~2d, we determined the coupling coefficients : $g_0=1.7$\AA$^{-1}$ and $g_1=220$\AA$^{-3}$, when $M=1.12 \times 10^6 m_e$.
For example, the real-time profile obtained from the displacement of $Q_{A_{u}}^{init}=40m$\AA~is presented in the inset of Fig.~5a. 
The amplitude for the induced vibration, denoted as $Q_{E_{u,x}}^{ind}$, is deduced from the time series $Q_{E_{u,x}}(t)$ and shown with respect to the various initial displacement of $Q_{A_u}^{init}$, as shown in Fig.~5a. 
The lattice dynamics, and the ensuing polarization variation, can be described with various choices of coordinates, such as bond angles as used in the previous literature.\cite{Fei2016} 
In the present work, in order to efficiently manifest the interaction between phonon modes, the displacements along the phonon modes are selected. 
The efficiency of this selection of coordinates can be observed from the molecular dynamics calculation results (Fig.~5a); 
the dynamical path obtained from the molecular dynamics is quite nicely described with the model Hamiltonian written in terms of these phonon coordinates.
A comparison with Fig.~5a and the Fig.~3a explains that the nonlinear coupling is negligible for small $Q_{A_u}$, but increases almost parabolically as $Q_{A_u}\geq 20m$\AA.

We now show that the coupling strength between the two phonons can be adjusted by an electrostatic gating: the hole doping.
We evaluate the same $Q_{E_{u,x}}^{ind}$, as introduced above, by $Q_{A_u}^{init}=80$m\AA~ with the change of Fermi level, as shown in Fig.~5b.
Overall, the increase in the hole concentration enhances the nonlinear phononic coupling strength.
The electronic origin for this variation with hole-doping can be analyzed with the electric part of Born effective charge: $Z_{\alpha }^\beta=V_{cell}\frac{dP_{\alpha}^{elec}}{dR_{I,\beta}}$.\cite{Gonze1997}
Assuming the adiabatic evolution of the electronic states along a specific phonon mode $Q$, one can obtain the $k$-resolved $Q$ mode effective charge ($\bold{\Omega}^{Q}(\bold{k})=-\frac{1}{e}\frac{d\bold{P}^{elec}(\bold{k})}{dQ}$) \cite {Cockayne2000} for the $n$-th band using a Kubo formula:\cite{Xiao2010,Min2014,Mele2002}
\begin{equation}
 \bold{\Omega}^Q_{n}(\bold{k})= 2\text{Im}\left[\sum_{m} 
\frac{\langle\psi_{n,\bold{k}}|\hat{\bold{v}}|\psi_{m,\bold{k}}\rangle \langle\psi_{m,\bold{k}}|\nabla _Q \hat{H}_{KS}|\psi_{n,\bold{k}}\rangle } {(\epsilon_{n,\bold{k}}-\epsilon_{m,\bold{k}})^2}\right] ,
 \end{equation}
where $\hat{\bold{v}}$ , $\nabla _Q \hat{H}_{KS}$ and $\epsilon$ are the velocity operator, the gradient of Kohn-Sham Hamiltonian with respect to atomic displacement of the phonon mode ($Q$), and the eigenvalue of Kohn-Sham state, respectively.
The electronic part of the mode effective charge can be obtained by integrating all occupied contributions over the Brillouin zone. 
For example, the $x$-component of the $Q$ mode effective charge is given by 
\begin{equation}
Z_{x}^Q=-eV_{cell}\sum_n \int_{BZ} d\bold{k}\int ^{\epsilon_F} _{-\infty} \hat{x} \cdot \bold{\Omega}^Q_{n} (\bold{k}) \delta(\epsilon_n(\bold{k})-\epsilon^{\prime})d\epsilon^{\prime}, 
\end{equation}
Figure~5c shows the $x$-component of the $A_u$ mode effective charge, which is sharply asymmetric against the sign change in $Q_{A_u}$.
As a result, we can deduce the fact that, for a light hole doping (in the range $-0.17 \leq \epsilon_F \leq 0$ in Fig.~5c), the in-plane polarization always increase as $\Delta P_x^{elec}=Z_{x}^{Q_{A_u}}{{Q_{A_{u}}}}>0$ for both positive and negative $Q_{A_u}$, which is in line with the explicit computation shown in the inset of Fig.~3a. 
On the other hand, the increasing nonlinear phonon coupling strength of the two modes with the hole doping, shown in Fig.~5b, can also be explained by the trend of the local charge distribution and mode effective charge on the hole doping. 
For example, for a given positive $Q_{A_u}=80m$\AA~as shown in Fig.~5c, the $A_u$ mode effective charge decreases with the hole doping, which implies the reduced amount of the electric polarization from $\Delta P_x^{elec}=Z_{x}^{Q_{A_u}}{Q_{A_{u}}}$ with a given $Q_{A_u}$.
This change of the in-plane polarization should be reflected as a variation in the electronic dipole moment.
This modified mode effective charge can be understood by Coulomb interaction of modified local charge distribution by hole doping.  The reduction of ionicity decreases the Coulomb attraction between Sn and Te from the local charge distribution at equilibrium geometry (depicted by $\delta$ in the inset of Fig.~5c), and thus leads to the increase of the repulsive restoring forces in the bond-compression regime, as depicted in the middle and bottom parts of Fig.~3c.
This intensified repulsion between Sn and Te results in the increase in-plane displacement, which eventually results in an enhanced responses of the $E_{u,x}$ amplitude with a given $Q_{A_u}$. 
The effect of doping on polar distortions has also discussed previously for bulk SnTe.\cite{Wang2020} We expect that the modified Coulomb interaction by doping not only alters the ground state configuration but modifies the lattice dynamics and the interaction strength between phonons for both bulk and monolayer SnTe.
Variations of mode effective charge and Born effective charge are discussed in Supplementary Note 3 and Supplementary Table 1 and 2.

Besides optical responses of this inversion-asymmetric insulators, as mentioned above, the carrier dynamics in the hole-doped material have been intriguing in terms of nonlinear Hall effect and second-order responses.\cite{Sodemann2015,Kim2019,Ma2019} 
The Berry curvature dipole have attracted intense interests for last couple of years.\cite{Sodemann2015,Ma2019}
The coupling between an external electric field and the Berry curvature dipole of the system can result in a second-order response current of $\bold{J}=\frac{e^3\tau}{2(1+i\omega\tau)}\hat{z} \times \bold{E}^* (\bold{\Lambda} \cdot \bold{E}) $, where $\Lambda$ is Berry curvature dipole ($\Lambda_a=\int_{\bold{k}}f^{occ}(\bold{k})\partial_a \Omega(\bold{k})$), $\hat{z}$ is the normal unit vector perpendicular to the plane made up of $D$ and $E$, and $f_{occ}$, $\tau$ are occupation of Bloch state and constant.\cite{Sodemann2015} 
As discussed previously, the Berry curvature dipole directly scales with the polarization that measures the amount of inversion-symmetry breaking of the lattice.\cite{Sodemann2015,Kim2019}
Here, we focus on the fact that the lattice distortion along the $A_u$ mode enhances the net in-plane polarization, and leads to an increase in the Berry curvature dipole. 
The static DFT calculation of the Berry curvature dipoles are presented in Fig.~5d, which confirms an almost monotonic increase of the Berry curvature dipole with the lattice displacement in $A_u$ mode eigenvector.
With the enhanced coupling strength, as shown in Fig.~5b, the nonlinear phononic interaction is to be more apparently revealed by the nonlinear Hall current and second harmonic generations of the Fermi level carriers of the hole-doped system, which is mediated by the Berry curvature dipole. 
These results indicate that second-order optical responses, such as nonlinear Hall effect and shift current, can be enhanced by nonlinear phonon interactions.

\section*{Discussion}
In summary, we demonstrated that nonlinear couplings between optical phonons in monolayer SnTe, a two-dimensional in-plane ferroelectric system, produce a polarization oscillations of the in-plane polarization in response to the pumped out-of-plane phonon.
We expect that this nonlinear phonon interaction can be generally observed from group-IV monochalcogenide monolayers which shares similar electronic and atomic structures. (Potential energy surfaces with respect lattice distortions of monolayer SnS are discussed in Supplementary Note 4 and Supplementary Figure 4.)
Through first-principles dynamical simulations, we revealed that a pumped $A_u$ phonon mode over nonlinear regime excites the $E_{u,x}$ modes, which in turn increases the in-plane electric polarization.
This charge oscillation originated from the nonlinear phonon coupling can be utilized for a dynamical control of ferroelectricities and optical response functions.
The coupling strength between the phonons is further enhanced by hole doping and thus the second order optical responses of the hole-doped carriers can also be pursued to evidence the nonlinear phononic coupling.   

\section*{Methods}
\subsection*{Computational details for first principle calculation}
To investigate the electronic structure and effect of phonons of monolayer SnTe, we performed DFT calculation using the Quantum Espresso package.\cite{Giannozzi2017}
To describe the electron-electron exchange correlation potential, Perdew-Burke-Ehrenof type generalized gradient approximation is employed.\cite{Perdew1996}
The full-relativistic norm-conserving pseudopotential is considered to describe spin-orbit interaction.
The plane wave basis set with 50 Ry energy cut-off is used to describe wavefunction under three-dimensional periodic boundary condition.
The lattice parameters are employed from the experimental observation with vacuum slab up to $20$\AA.\cite{Chang2016,Kim2019}
The Bloch states are considered with k-point sampling up to $50 \times 50 \times 1$, and convergence is checked up to $80 \times 80 \times 1$ k-point mesh.
The electric polarization of the 2D periodic system is evaluated following equation: $P=P_e +\frac{e}{\Omega}\Sigma_i Z_i \bold{R}_i$, where $P_e=\frac{ie}{(2\pi)^3}\Sigma_i\int_{\bold{k}}<u_{i,\bold{k}}|\nabla_{\bold{k}}|u_{i,\bold{k}}>$ is the electric polarization of the electrons, and $Z_i$ and $\bold{R}_i$ are the ionic charge and the position of the $i$-th atom, respectively.\cite{King-Smith1993}
For the BO molecular dynamics simulation, Verlet algorithm is employed with time step $dt=0.48$~fs. We also checked on that the total system energy is conserved within $0.1$~meV during molecular dynamics time of $2.1$~ps.
Real-time propagation of Bloch state were performed by using our in-house computation packages.\cite{Shin2016}
Detailed techniques related to the time-integration of the time-dependent DFT and Kohn-Sham equations are well described in our previous works.\cite{Shin2016,Shin2019}

 \subsection*{Data availability statement} All relevant data is included in the main manuscript. All the data generated and analysed during this study are available from the corresponding authors upon reasonable request.

 \subsection*{Acknowledgements} We are grateful for helpful discussions with Hosub Jin and Jungwoo Kim. We further acknowledge financial support from the European Research Council (ERC-2015-AdG-694097), the Clusters of Excellence “Advanced Imaging of Matter” (AIM, EXC 2056, ID 390715994), Grupos Consolidados (IT1249-19), and SFB925. D.S. acknowledges the support from National Research Foundation of Korea (NRF-2019R1A6A3A03031296). N.P. was supported by National Research Foundation of Korea (NRF-2019R1A2C2089332). The Flatiron Institute is a division of the Simons Foundation. 
 
 \subsection*{Competing Interests} The authors declare that they have no competing financial interests.
 
 \subsection*{Author contributions}  D.S. performed the calculations; D.S., S.S., H.H. and U.D.G. analyzed the data; D.S., N.P. and A.R. wrote the paper. All authors discussed and analyzed the results and contributed and commented on the manuscript.
 
 \subsection*{Correspondence} Correspondence and requests for materials should be addressed to N.P. (noejung@unist.ac.kr) and A.R. (angel.rubio@mpsd.mpg.de).
 \subsection*{Supplementary information} Supplementary information is available for this paper at https://doi.org/.

\section*{Figure Legends}

\begin{figure}
\centering
\includegraphics[width=0.80\textwidth]{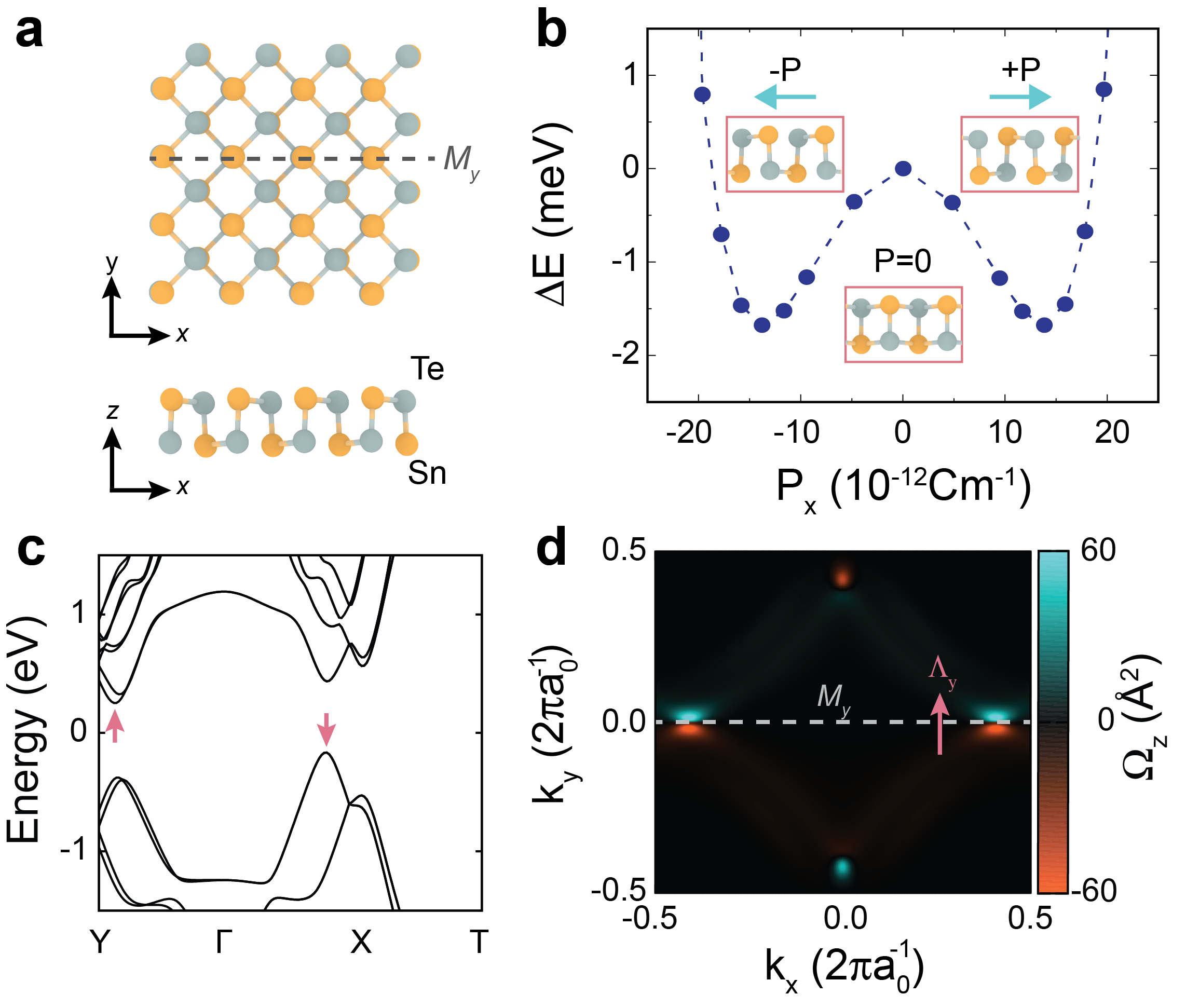}
\caption{ {\bf Atomic geometry and electronic structure of monolayer SnTe.}
{\bf a} Atomic geometry of monolayer SnTe with top view (upper diagram) and side view (lower diagram). 
{\bf b} Potential energy surface of monolayer SnTe with respect to the in-plane electric polarization. 
{\bf c} The electronic band structure and {\bf d} Berry curvature distribution of the monolayer SnTe. 
In {\bf a} and {\bf b}, orange and gray balls indicate Sn and Te atoms, respectively. 
In {\bf b}, green arrows indicate the electric polarization.
In {\bf c}, up and down arrows indicate valence band maximum and conduction band minimum states, respectively.
Dashed horizontal lines in {\bf a} and {\bf d} indicates the mirror plane ($M_y$). 
The Berry curvature dipole for the hole-doped case is denoted by $\Lambda_y$ in {\bf d}. 
}
\end{figure}

\begin{figure}
\centering
\includegraphics[width=0.70\textwidth]{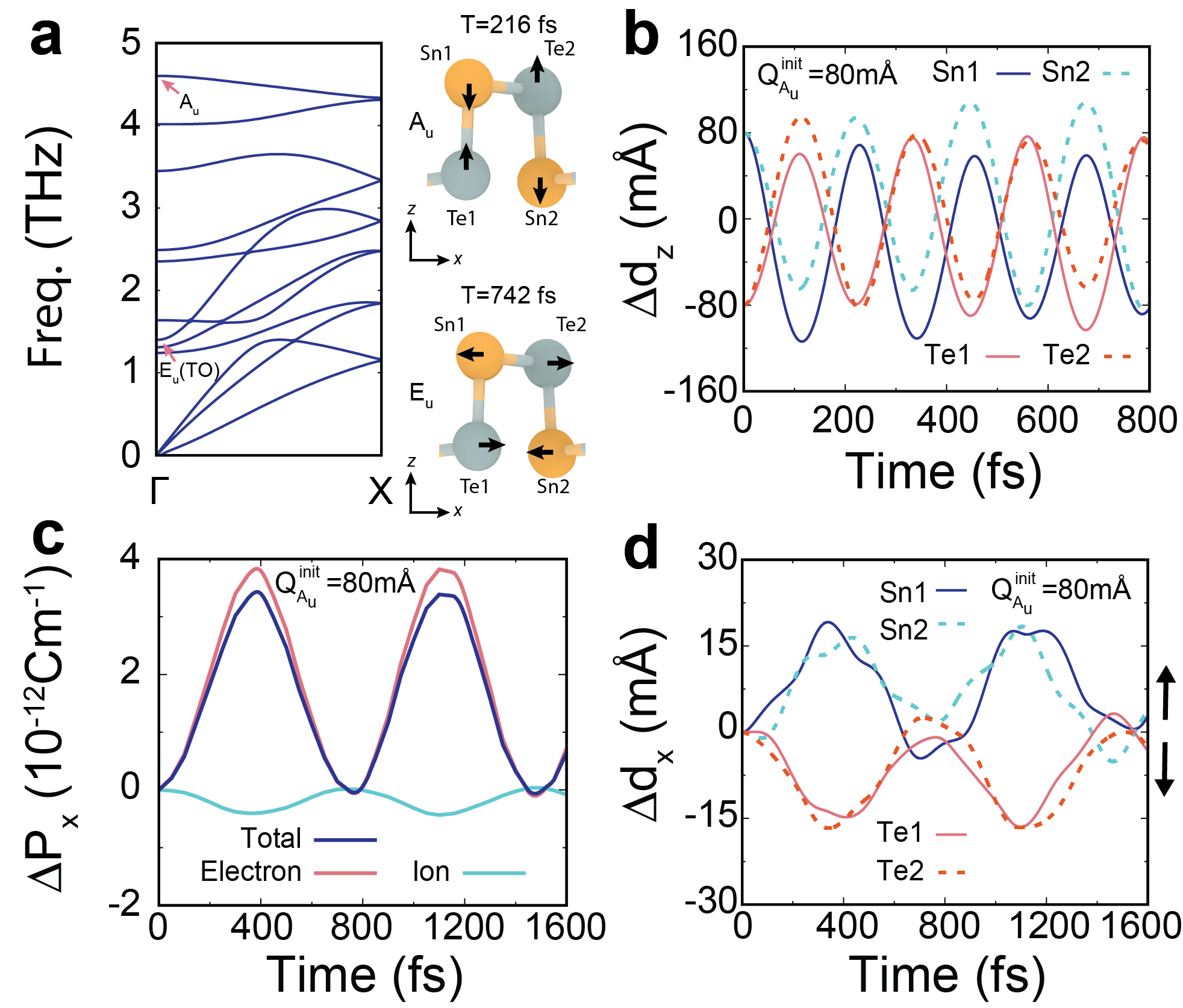}
\caption{ {\bf Nonlinear phonon interaction in monolayer SnTe.}
{\bf a} Phonon dispersion and eigenvectors of the IR-active $A_u$ and $E_{u,x}$ phonon modes. 
{\bf b} Time-profile of atomic motions in the $z$-direction during the BO dynamics initiated with $Q_{A_u}^{init}=80m$\AA.
{\bf c} Time-profile of the in-plane electric polarization induced by initially pumped $A_u$ phonon ($Q_{A_u}^{init}=80m$\AA). 
{\bf d} The same atomic motions in the $x$-direction.
}
\end{figure}

\begin{figure}
\centering
\includegraphics[width=0.80\textwidth]{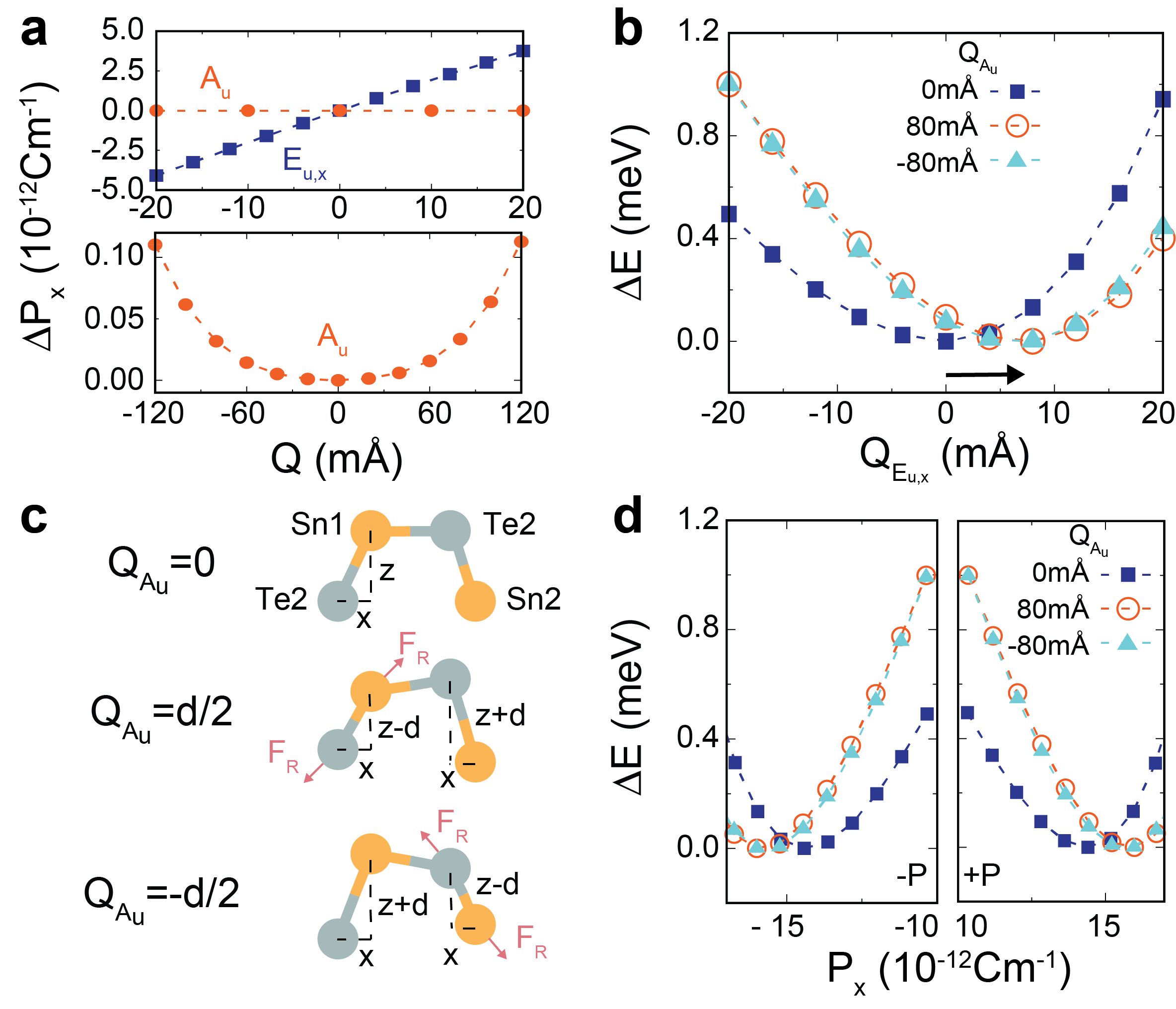}
\caption{ {\bf Enhanced electric polarization by nonlinear phonon interaction.}
{\bf a} Variation of the in-plane polarization by displacements either in the $A_u$ and the $E_{u,x}$ phonon mode (top) and the same variation with the $A_u$ mode in the extended window (bottom). 
{\bf b} The potential energy change for additional displacements along to the $E_{u,x}$ phonon mode on top of a given fixed displacement along the $A_u$ phonon mode ($Q_{A_u}$).
{\bf c} Schematic atomic geometry of SnTe without (upper) and with (middle and below) displacements along the $A_u$ phonon mode.
{\bf d} The potential energy surface of the $E_{u,x}$ mode with respect to the in-plane electric polarization.
In {\bf b} and {\bf d}, the energy is plotted with reference to its minimum at a given $Q_{A_u}$: $\Delta E= E[Q_{A_u},Q_{E_{u,x}}]-E^{min}_{Q_{A_u}}[Q_{E_{u,x}}]$.
}
\end{figure}

\begin{figure}
\centering
\includegraphics[width=0.70\textwidth]{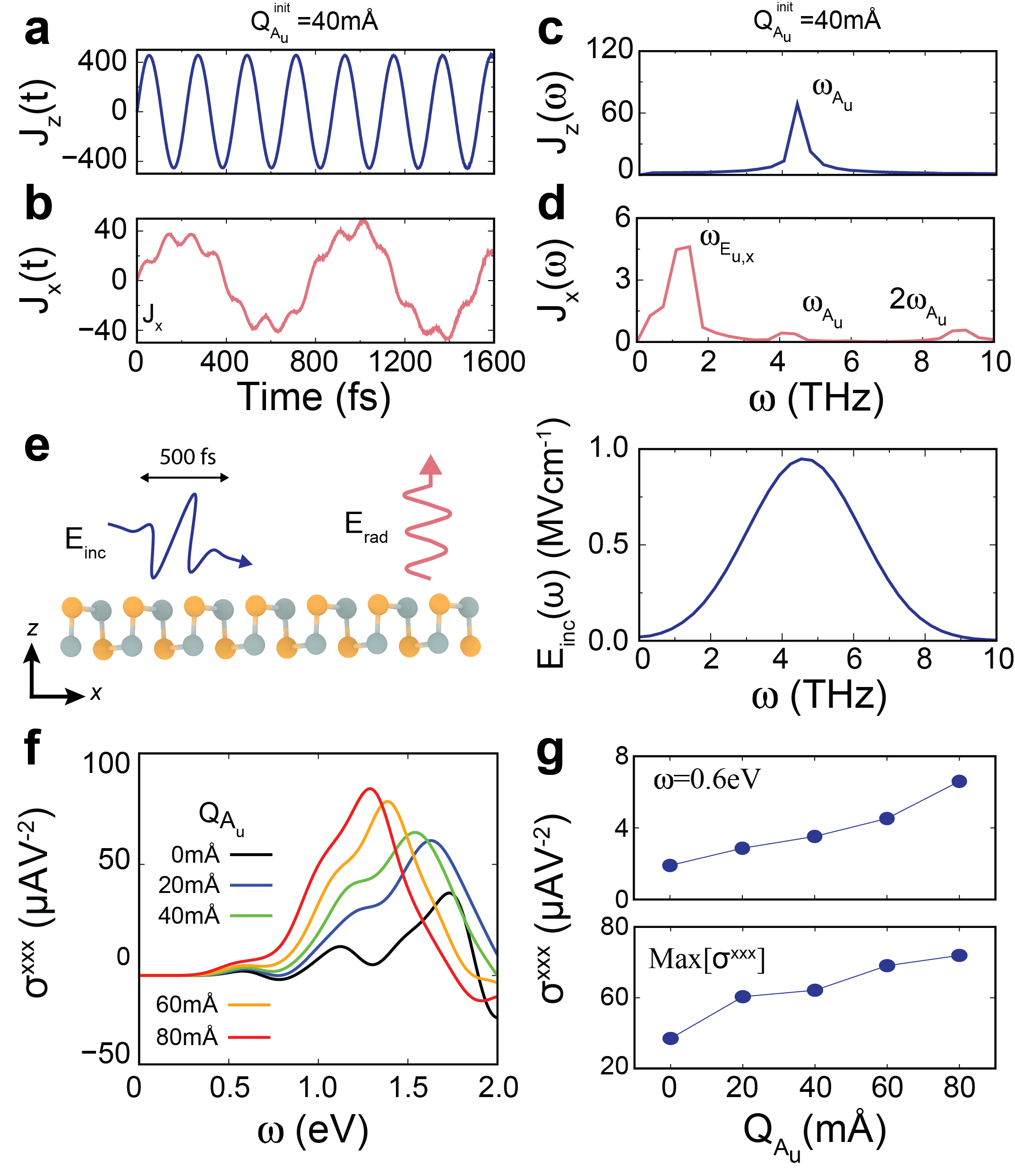}
\caption{ {\bf Induced current oscillation and photovoltaic effect with nonlinear phonon interaction.}
Real-time profile of the calculated electronic current {\bf a} $J_z(t)$ and {\bf b} $J_x(t)$ induced by initial atomic displacement toward the $A_u$ phonon mode ($Q_{A_u}^{init}$).
Corresponding Fourier components for {\bf a} and {\bf b} are given in {\bf c} and {\bf d}, respectively. 
{\bf e} Schematic illustration of the incident laser polarized along $z$-direction (blue) and the radiation field polarized along $x$-direction (pink). Right panel shows the band width of the incident pulse peak at the frequency of the $A_u$ phonon mode.
{\bf f} Shift current response function for fixed geometry with given $A_{u}$ phonon displacement.
{\bf g} The shift current at $\omega=0.6eV$ (upper) and the maximum value (below).
}
\end{figure}

\begin{figure}
\centering
\includegraphics[width=0.70\textwidth]{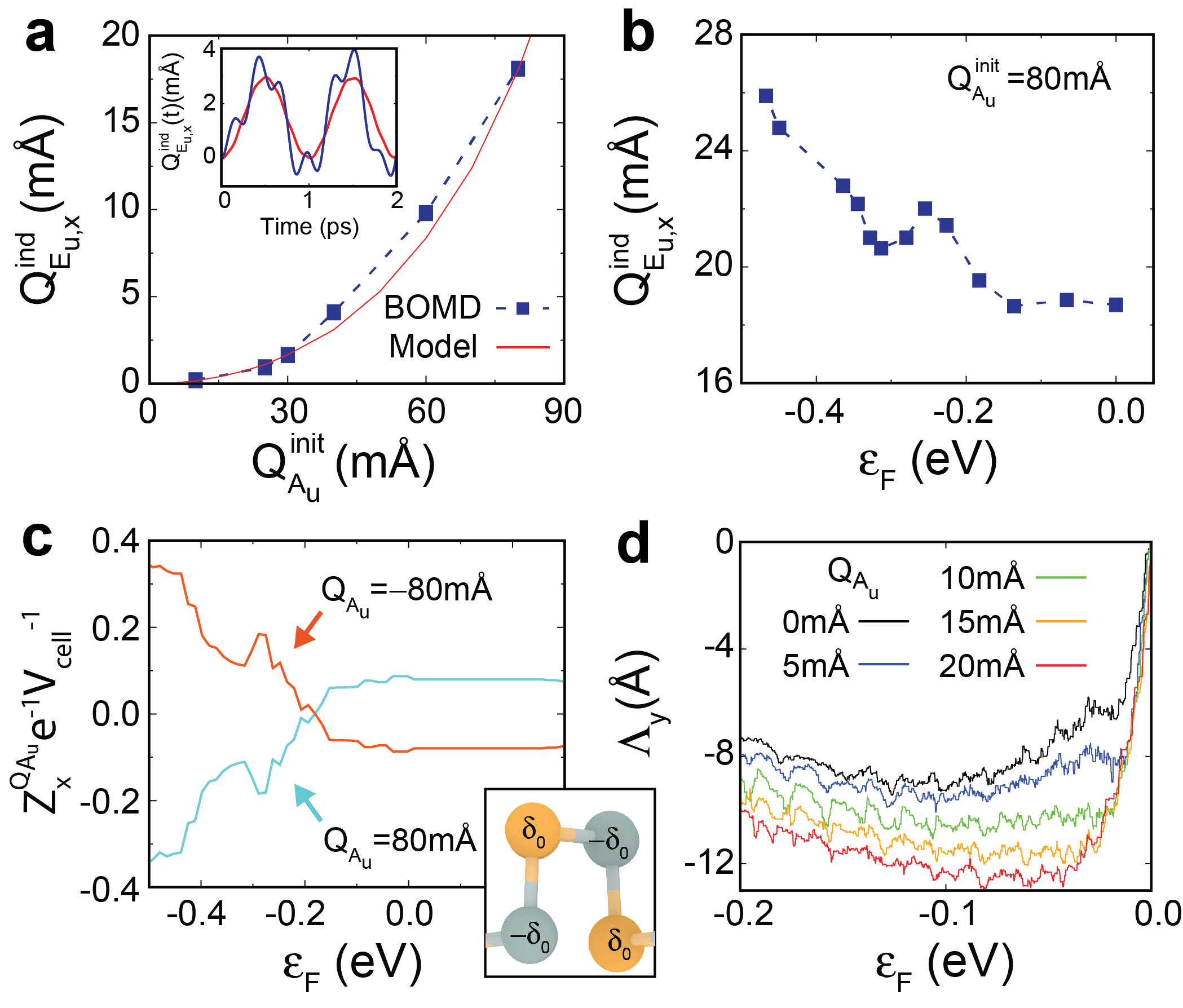}
\caption{ {\bf Doping property of nonlinear phonon interaction and enhanced nonlinear Hall effect.}
{\bf a} Induced amplitude of the $E_{u,x}$ phonon mode with respect to initial atomic displacement in $A_u$ phonon mode ($Q^{init}_{A_u}$). 
{\bf b} The same as {\bf a} with $Q_{A_u}^{init}=80m$\AA~ with various levels of hole doping. 
{\bf c} Variation in the electronic part of the $A_u$ mode effective charge.
{\bf d} The Berry curvature dipole of the monolayer SnTe with a fixed displacement along $A_u$ phonon eigenvector ($Q_{A_u}$) with respect the Fermi level shift. 
Inset of {\bf a} indicates the real-time profile of $Q_{E_{u,x}}(t)$ induced by initial $Q_{A_{u}}^{init}=40m$\AA.
Inset of {\bf c} schematically sketches the net charge of the atoms to characterize the ionicity at equilibrium points.
}
\end{figure}

\end{document}